\documentclass[twocolumn]{revtex4}

\input psfig.sty

\usepackage{graphicx}
\usepackage{dcolumn}
\usepackage{bm}

\begin{document}

\title{Exact solutions of Brans-Dicke cosmology with decaying vacuum density}

\author{A. E. Montenegro Jr.$^1$}

\author{S. Carneiro$^{1,2}$}

\affiliation{$^1$Instituto de F\'{\i}sica, Universidade Federal da
Bahia, 40210-340, Salvador, BA, Brazil \\ $^2$International Centre
for Theoretical Physics, Trieste, Italy\footnote{Associate Member}}

\begin{abstract}
We investigate cosmological solutions of Brans-Dicke theory with
both the vacuum energy density and the gravitational constant
decaying linearly with the Hubble parameter. A particular class of
them, with constant deceleration factor, sheds light on the
cosmological constant problems, leading to a presently small
vacuum term, and to a constant ratio between the vacuum and matter
energy densities. By fixing the only free parameter of these
solutions, we obtain cosmological parameters in accordance with
observations of both the relative matter density and the universe
age. In addition, we have three other solutions, with Brans-Dicke
parameter $\omega=-1$ and negative cosmological term, two of them
with a future singularity of big-rip type. Although interesting
from the theoretical point of view, two of them are not in
agreement with the observed universe. The third one leads, in the
limit of large times, to a constant relative matter density, being
also a possible solution to the cosmic coincidence problem.
\end{abstract}

\maketitle

\section{Introduction}

Recent observations suggest that the total energy density in the
universe is greater than the (barionic + dark) matter density. On
one hand, dynamical estimations lead to a ratio between the matter
and critical densities around one third \cite{calb}. On the other
hand, measurements of anisotropies in the microwave background
radiation indicate that our universe is spatially flat
\cite{bach}, suggesting the existence of another, unknown form of
energy, usually called dark energy. Its presence is also
corroborated by age estimations \cite{age} and by the observed
distance-redshift relation for supernovas Ia, which suggests that,
in the present phase of universe evolution, the deceleration
parameter $q$ is negative, and therefore the universe performs an
accelerated expansion \cite{SnIa}.

Several models have been proposed in order to explain those
observational data. There are some candidates for dark energy, as,
for example, the cosmological constant, the so called quintessence,
or the generalized Chaplygin gas. Among them, the simplest and
oldest one is the cosmological constant, also associated to the
energy density of vacuum.

We can contextualize the study of dark energy in different
gravitational theories. For instance, the theory of General
Relativity, where the Einstein field equations are used, and where
the gravitation constant $G$ is a universal constant. Another one is
the Brans-Dicke theory \cite{weinberg,brans}, a scalar-tensor theory
in which the gravitational constant is a function of space-time, and
where a new parameter, $\omega$, is introduced. More recently the
interest on this kind of theory was renewed, owing to its
association with superstrings theories, extra-dimensional theories
and models with inflation or accelerated expansion
\cite{pavon}-\cite{barrow}.

It is generally assumed that General Relativity is recovered in the
limit $\omega\rightarrow+\infty$ (despite the existence of
Brans-Dicke solutions for which this is not true \cite{faraoni}).
Astronomical observations in the realm of Solar System impose a very
high inferior limit for $\omega$. Nevertheless, such a result
corresponds to the weak field limit, and applies only in the
simplest case of constant $\omega$. Therefore, it is possible that
General Relativity is not adequate to describe the universe at early
times, or needs corrections in the cosmological limit.

In this article we consider the Brans-Dicke theory and associate to
dark energy the equation of state of vacuum. We investigate models
in which the vacuum energy density decreases with the universe
expansion, a hypothesis that has been considered as a possible
solution to the cosmological constant problem, that is, to the
question of why the presently observed value of $\Lambda$ is about
$120$ orders of magnitude below the value predicted by quantum field
theories \cite{sahni,peebles}.

Our goal is to find solutions of Brans-Dicke theory which satisfy a
particular variation law for $G$. We shall use the Eddington-Dirac
relation, based on the large number coincidence, $G\approx
H/m_\pi^3$, where $H=\dot{a}/a$ is the Hubble parameter and $m_\pi$
is the pion mass \cite{saulo2,saulo1}. We will then take
$G=H/8\pi\lambda$, where $\lambda$ has the order of $m_\pi^3$. In
addition, as usual, we will relate the Brans-Dicke scalar field to
the gravitational constant through $\phi=G_0/G$, where $G_0$ is a
positive constant of the order of unity.

Together with that variation law for $G$, we shall consider two
different ansatzen. The first one is given by $\rho=3\alpha H^2/8\pi
G$, where $\rho=\rho_m+\rho_\Lambda$ is the total density, and
$\alpha$ is an adimensional constant of the order of unity. This
ansatz is suggested by observations, which show that
$\rho_m\approx\rho_c/3$, where $\rho_c=3H^2/8\pi G$ is the critical
density. On the other hand, we know that $\rho_\Lambda$ has, at
most, the same order of magnitude as $\rho_m$, otherwise its
presence would be more evident. This ansatz was already considered
in \cite{saulo3,agostinho}.

The second ansatz will be given by $\Lambda=\beta H^2$, where
$\beta$ is a constant of the order of unity. We are, in this case,
inferring a variation law for the cosmological term, which has
already been considered in the literature on the basis of different
arguments \cite{saulo1}-\cite{aldro} (for other variation laws for
the cosmological term, see, for instance, \cite{wu}-\cite{prd}). We
will show that this ansatz leads to a set of solutions larger than
the first one, containing its solutions as a particular case.

We will look for solutions for recent times, that is, we shall
consider a spatially flat ($k$ = 0) Friedmann-Robertson-Walker
space-time, filled with a perfect fluid whose matter component is
pressureless ($p_m = 0$). For the cosmological term, we will take
the equation of state of vacuum, $p_\Lambda=-\rho_\Lambda$.

\section{Solutions with varying cosmological term}

\subsection{The first ansatz}

Taking $p_m=0$, $k=0$ and $p_\Lambda=-\rho_\Lambda$, the Brans-Dicke
equations \cite{weinberg} can be written as
\begin{equation}\label{60}
\frac{d(\dot{\phi}
a^3)}{dt}=\frac{8\pi}{3+2\omega}\left(\rho+3\rho_\Lambda\right)a^3,
\end{equation}
\begin{equation}\label{61}
\dot{\rho}=-3H\rho_m,
\end{equation}
\begin{equation}\label{62}
H^2=\frac{8\pi\rho}{3\phi}-\frac{\dot\phi}{\phi}H+\frac{\omega}{6}\frac{\dot\phi^2}{\phi^2}.
\end{equation}

We then have a system of three ordinary differential equations, with
four unknown functions of time: $a$, $\rho_m$, $\rho_{\Lambda}$ and
$\phi$. The system becomes solvable if we add the Eddington-Dirac
relation, $G=H/8\pi\lambda$, which relates $a$ and $\phi$. In order
to restrict our class of solutions, we will take in addition our
first ansatz, given by $\rho=3\alpha H^2/8\pi G$. In this way, we
obtain
\begin{equation}\label{63}
\rho=3\alpha\lambda H,
\end{equation}
\begin{equation}\label{64}
\phi=\frac{8\pi\lambda G_0}{H},
\end{equation}
and
\begin{equation}\label{65}
\dot{\phi}=8\pi\lambda G_0(1+q),
\end{equation}
where $q=-\ddot{a}a/\dot{a}^2$ is the deceleration factor.

With the help of equations (\ref{63})-(\ref{65}),  we can rewrite
(\ref{60})-(\ref{62}) in the form
\begin{equation}\label{66}
(3+2\omega)\lambda G_0[\dot{q}+(1+q)3H]=3\alpha\lambda
H+3\rho_\Lambda,
\end{equation}
\begin{equation}\label{67}
\rho_m=\alpha\lambda H(1+q),
\end{equation}
\begin{equation}\label{68}
\frac{\alpha}{G_0}=2+q-\frac{\omega}{6}(1+q)^2.
\end{equation}

Equation (\ref{68}) tell us that $q$ is constant, since $\alpha$,
$G_0$ and $\omega$ also are. Therefore, $\dot{q}=0$, and equation
(\ref{66}) reduces to
\begin{equation}\label{69}
(3+2\omega)\lambda G_0(1+q)H=\alpha\lambda H+\rho_\Lambda.
\end{equation}

By using (\ref{63}) and (\ref{67}), we can obtain the vacuum
density,
\begin{equation}\label{70}
\rho_\Lambda=\alpha\lambda H(2-q).
\end{equation}
Substituting $\lambda=H/8\pi G$, we obtain
\[\rho_\Lambda=\frac{\alpha(2-q)H^2}{8\pi G}.\]
Since $\rho_\Lambda=\Lambda/8\pi G$, we conclude that
$\Lambda=\alpha(2-q)H^2$, which suggests our second ansatz, with
$\beta=\alpha(2-q)$, to be used later.

Leading (\ref{63}) and (\ref{67}) into (\ref{61}), one obtains
\[\frac{1}{H}=(1+q)t+C,\] where $C$ is an integration constant.
Let us take $C=0$, in such a way that $H\rightarrow\infty$ for
$t\rightarrow0$. We then have
\begin{equation}\label{71}
H=\frac{1}{1+q}\frac{1}{t}.
\end{equation}

Substituting $\dot{a}/a$ for $H$ in (\ref{71}), we also have
\begin{equation}\label{72}
a=At^{\frac{1}{1+q}},
\end{equation}
where $A$ is another integration constant.

The relative density of matter, defined with respect to the critical
density, can be obtained by using $G=H/8\pi\lambda$. Then,
$\rho_c=3\lambda H$, and, using (\ref{67}), one obtains
\begin{equation}\label{73}
\Omega_m=\frac{\rho_m}{\rho_c}=\frac{\alpha(1+q)}{3}.
\end{equation}

Substituting $\rho_\Lambda$ from (\ref{70}) into (\ref{69}), we
have as well
\begin{equation}\label{76}
\frac{\alpha}{G_0}=\frac{(3+2\omega)(1+q)}{3-q}.
\end{equation}
Comparing $\alpha/G_0$ given by equations (\ref{68}) and (\ref{76}),
one can derive a relation between $\omega$ and $q$, given by
\begin{equation}\label{77}
(3+2\omega)(1+q)=\left[2+q-\frac{\omega}{6}(1+q)^2\right](3-q).
\end{equation}
Eliminating $\omega$ from equations (\ref{68}) and (\ref{76}), we
can also obtain $\alpha/G_0$ as a function of $q$ only,
\begin{equation}\label{78}
\frac{\alpha}{G_0}=\frac{12(2+q)+3(1+q)^2}{(1+q)(3-q)+12}.
\end{equation}

With these relations, it is easy to derive some results to be
compared with current observations. For example, if $q = 0$, from
equations (\ref{77}) and (\ref{78}) we obtain $\omega=6/5$ and
$\alpha/G_0=9/5$. From (\ref{71}) we have $Ht=1$. From equation
(\ref{72}) it follows that $a=At$. And, from (\ref{73}), one has
$\Omega_m/\alpha=1/3$. Since $\alpha\approx1$, we can see that
$\Omega_m\approx1/3$, in agreement with astronomical estimations
\cite{calb}. The age parameter $Ht$ is also in good accordance
with globular clusters observations \cite{age}.

If, on the other hand, we would take $q = -1$, we would obtain,
instead of equation (\ref{71}), the result $H=$ constant, that is,
the de Sitter universe, with $\rho_m=0$ and a constant
$\rho_\Lambda$. Note, however, that $q = -1$ does not satisfy
equation (\ref{77}), that is, the de Sitter universe is not
solution of Brans-Dicke equations for this ansatz.

\subsection{The second ansatz}

Taking now our second ansatz, $\Lambda=\beta H^2$ and $G=H/8\pi
\lambda$, and reminding that $\rho_\Lambda=\Lambda/8\pi G$, we
obtain
\begin{equation}\label{82}
\rho_\Lambda=\beta\lambda H.
\end{equation}
Furthermore, as well as in the first ansatz, we have
\begin{equation}\label{83}
\phi=\frac{8\pi\lambda G_0}{H}
\end{equation}
and
\begin{equation}\label{84}
\dot{\phi}=8\pi\lambda G_0(1+q).
\end{equation}

With the help of (\ref{82})-(\ref{84}), we can put
(\ref{60})-(\ref{62}) in the form
\begin{equation}\label{85}
(3+2\omega)\lambda G_0[\dot{q}+3H(1+q)]=\rho+3\beta \lambda H,
\end{equation}
\begin{equation}\label{86}
\dot{\rho}+3H\rho-3\beta \lambda H^2=0,
\end{equation}
\begin{equation}\label{87}
\rho=3\lambda G_0H\left[2+q-\frac{\omega}{6}(1+q)^2\right].
\end{equation}
Here we have a solvable system, with three differential equations
for three unknown functions, $H$, $\rho$ and $q$. By finding
$\rho$ one can, using (\ref{82}), determine $\rho_{\Lambda}$ and
$\rho_m$.

Leading $\rho$ given by (\ref{87}) into (\ref{85}), we derive
\begin{equation}\label{88}
\frac{\beta}{G_0}=\frac{(3+2\omega)[\dot{q}+3H(1+q)]-3H[2+q-\frac{\omega}{6}(1+q)^2]}{3H}.
\end{equation}
In this way, since $\beta/G_0$ is constant, there are two
possibilities: either $\dot{q}=0$, in which case we have a simple
relation between $q$ and $\beta/G_0$, or (\ref{88}) is an evolution
equation, with $\dot{q} \neq 0$.

\subsubsection{The case $\dot{q}=0$}

In this case, equation (\ref{88}) becomes
\begin{equation}\label{89}
\frac{\beta}{G_0}=(3+2\omega)(1+q)-\left[2+q-\frac{\omega}{6}(1+q)^2\right].
\end{equation}

Using (\ref{87}) into (\ref{86}), one obtains, by integration,
\begin{eqnarray}
\frac{1}{H}=\frac{3G_0[2+q-
\frac{\omega}{6}(1+q)^2]-\beta}{G_0[2+q-\frac{\omega}{6}(1+q)^2]}t+C
\nonumber.
\end{eqnarray}

Let us choose $C=0$, such that $H\rightarrow\infty$ for
$t\rightarrow0$. So we have
\begin{equation}\label{90}
H=\frac{n}{t},
\end{equation}
where we have defined
\begin{equation}\label{91}
n=\frac{G_0[2+q-\frac{\omega}{6}(1+q)^2]}{3G_0[2+q-\frac{\omega}{6}(1+q)^2]-\beta}.
\end{equation}

Substituting $\dot{a}/a$ for $H$ in equation (\ref{90}), we find
\begin{equation}\label{92}
a=At^n,
\end{equation}
where $A$ is an integration constant. On the other hand, from
(\ref{92}) we can obtain $q=(1-n)/n$, or
\begin{equation}\label{93}
n=\frac{1}{1+q}.
\end{equation}

Leading $n$ from equation (\ref{91}) into (\ref{93}), one obtains
\begin{equation}\label{96}
\frac{\beta}{G_0}=\left[2+q-\frac{\omega}{6}(1+q)^2\right](2-q),
\end{equation}
and we can write
\begin{equation}\label{97}
\rho_\Lambda=\beta\lambda H=\lambda
G_0H\left[2+q-\frac{\omega}{6}(1+q)^2\right](2-q).
\end{equation}

By using (\ref{87}) and (\ref{97}), we also have
\begin{equation}\label{98}
\rho_m=\rho-\rho_\Lambda=\lambda
G_0H\left[2+q-\frac{\omega}{6}(1+q)^2\right](1+q).
\end{equation}

We can also obtain the relative density of matter from equations
(\ref{96}) and (\ref{98}). Reminding that $\rho_c=3\lambda H$, we
have
\begin{equation}\label{100}
\Omega_m=\frac{\rho_m}{\rho_c}=\frac{\beta}{3}\left(\frac{1+q}{2-q}\right).
\end{equation}

Eliminating $\omega$ from (\ref{89}) and (\ref{96}), one can obtain
$\beta/G_0$ as a function of $q$ only,
\begin{equation}\label{104}
\frac{\beta}{G_0}=\frac{12(2+q)+3(1+q)^2}{(1+q)(3-q)+12}\;(2-q).
\end{equation}
On the other hand, comparing $\beta/G_0$ given by (\ref{89}) with
that given by (\ref{96}), we obtain the same relation between
$\omega$ and $q$ we have obtained with the first ansatz, equation
(\ref{77}).

This is not a mere coincidence. If we compare $\rho$ given by
(\ref{87}) with that obtained from equation (\ref{63}) of the
first ansatz, we obtain the equation (\ref{68}) of the first
ansatz. Then, (\ref{97}) can be reduced to equation (\ref{70}) of
the first ansatz. Equation (\ref{98}), on the other hand, is
reduced to equation (\ref{67}) of the first ansatz.

Equation (\ref{96}) can be put in the form
\begin{equation}\label{108}
\beta=\alpha (2-q),
\end{equation}
already anticipated in the first ansatz. By using it, it is
possible to verify that equations (\ref{100}) and (\ref{104}) are
the same as (\ref{73}) and (\ref{78}) of the first ansatz.
Finally, one can also verify, with the help of (\ref{93}), that
equations (\ref{90}) and (\ref{92}) are identical to equations
(\ref{71}) and (\ref{72}) of the first ansatz, respectively.

We thus conclude that, in the case of a constant $q$, the two
ansatzen are equivalent.

\subsubsection{The case $\dot{q} \neq 0$}

In the differential equation (\ref{88}), substituting $\dot{a}/a$
for $H$ and separating the variables, we obtain
\[\frac{dq}{\frac{\omega}{6}\left[\left(\frac{G_0(6\omega+6)^2+6\omega(G_0+\beta)}
{G_0\omega^2}\right)-\left(q+\frac{6+7\omega}{\omega}\right)^2\right]}=\frac{3}{3+2\omega}
\frac{da}{a}.\] We will initially analyze the case in which the
quantity
\[\kappa^2=\frac{G_0(6\omega+6)^2+6\omega(G_0+\beta)}{G_0\omega^2}\]
is positive. In sections \emph{2.3} and \emph{2.4} we shall
analyze the cases in which it is negative or zero, respectively.

With $\kappa^2>0$, let us integrate the above equation by doing
\[z=q+\frac{6+7\omega}{\omega}.\] Then, we obtain
\[\frac{6}{\omega}\int\frac{dz}{\kappa^2-z^2}=\frac{3}{3+2\omega}
\int\frac{da}{a}.\] Its solution is given by
\[a=A\left|\frac{\kappa+q+\frac{6+7\omega}
{\omega}}{\kappa-q-\frac{6+7\omega}{\omega}}\right|^{\frac{3+2\omega}{\omega
\kappa}},\] where $A$ is an integration constant.

By defining $B=\kappa+(6+7\omega)/\omega$,
$C=\kappa-(6+7\omega)/\omega$ and $D=(3+2\omega)/\omega\kappa$, we
have
\begin{equation}\label{109}
a=A\left|\frac{B+q}{C-q}\right|^D.
\end{equation}
Then,
\begin{equation}\label{110}
 \frac{B+q}{C-q}<0 \Rightarrow a=A\left(\frac{B+q}{q-C}\right)^D,
\end{equation}
while
\begin{equation}\label{111}
\frac{B+q}{C-q}>0 \Rightarrow a=A\left(\frac{B+q}{C-q}\right)^D.
\end{equation}
Let us solve equations (\ref{110}) and (\ref{111}), in order to
find the functions $q$, $H$ and $\rho$.

\vspace{8 mm}

{\emph{2.1 - The solution of equation (\ref{110})}

\vspace{5 mm}

Introducing $x=a/A$ and inverting equation (\ref{110}), we obtain
\begin{equation}\label{112}
q=\frac{B+Cx^{\frac{1}{D}}}{x^{\frac{1}{D}} -1}.
\end{equation}
With the definition of $q$, it becomes
\[\dot{x}^2(B+Cx^{\frac{1}{D}})+x\ddot{x}(x^{\frac{1}{D}}-1)=0.\]

By taking $y=\dot{x}$ and $y'=dy/dx$, one has
\[y(B+Cx^{\frac{1}{D}})+y'x(x^{\frac{1}{D}}-1)=0,\]
which solution is
\[y(x)=\frac{C_1x^B}{(x^{\frac{1}{D}}-1)^{D(B+C)}},\]
where $C_1$ is an integration constant.

It is easy to see that $H=y/x$, leading to
\begin{equation}\label{113}
H(x)=\frac{C_1x^{B-1}}{(x^{\frac{1}{D}}-1)^{D(B+C)}}.
\end{equation}
On the other hand, we have $dt=dx/y$, and so
\begin{equation}\label{114}
t=\int \frac{(x^{\frac{1}{D}}-1)^{D(B+C)}}{C_1x^B}\;dx.
\end{equation}

Equations (\ref{112})-(\ref{114}) are solutions of the Brans-Dicke
equations (\ref{85}) and (\ref{87}). Let us now verify whether they
satisfy the third Brans-Dicke equation, (\ref{86}). Introducing
$\rho'=d\rho/dx$, and using $\rho$ given by (\ref{87}), equation
(\ref{86}) becomes
\[3\lambda
G_0\frac{dH}{dx}\left[2+q-\frac{\omega}{6}(1+q)^2\right]\]
\[+\;3\lambda G_0H\frac{dq}{dx}\left[\frac{3-\omega(1+q)}{3}\right]\]
\[+\,\frac{3}{x}\left\{3\lambda
G_0H\left[2+q-\frac{\omega}{6}(1+q)^2\right]-\beta \lambda
H\right\}=0.\] By using (\ref{112}) and (\ref{113}), we obtain
\[\frac{\beta}{G_0}=-\frac{(C+1)x^{\frac{1}{D}}+(B-1)}{x^{\frac{1}{D}}-1}
\left[2+q-\frac{\omega}{6}(1+q)^2\right]\]
\[-\,\frac{(B+C)x^{\frac{1}{D}}}{D(x^{\frac{1}{D}}-1)^2}\left[\frac{3-\omega(1+q)}{3}
\right]+3\left[2+q-\frac{\omega}{6}(1+q)^2\right].\] Substituting
in this equation $x$ given by (\ref{110}), we then have
\[\frac{\beta}{G_0}=4-\frac{\omega}{3}-\frac{BC(\omega-3)}{3D(B+C)}\]
\[+\left[\frac{(B-C)
(\omega-3)-BC\omega}{3D(B+C)}-\frac{\omega}{2}\right]q\]
\begin{equation}\label{115}
+\left[\frac{\omega(B-C)+\omega-3}{3D(B+C)}-1\right]q^2+\left[\frac{\omega}
{3D(B+C)}+\frac{\omega}{6}\right]q^3.
\end{equation}

In this equation, $\omega$, $\beta/G_0$, $B$, $C$ and $D$ are
constants. Therefore, for a varying $q$, the coefficients of $q$,
$q^2$ and $q^3$ must be identically zero, simultaneously. This is
only possible for $\omega=-1$ and $\beta/G_0=-3$. As $G_0$ is
positive, we conclude that $\beta$ is negative, that is, the
cosmological term is negative.

We then have $B=\sqrt{12}+1$, $C=\sqrt{12}-1$ and $D=-1/\sqrt{12}$.
In this way, equations (\ref{112})-(\ref{114}) may be written as
\begin{equation}\label{117}
q=\frac{\sqrt{12}+1+(\sqrt{12}-1)x^{-\sqrt{12}}}{x^{-\sqrt{12}}-1},
\end{equation}
\begin{equation}\label{118}
H=C_1 x^{\sqrt{12}}\left(x^{-\sqrt{12}}-1\right)^2,
\end{equation}
\begin{equation} \label{119}
C_1t=\frac{1}{\sqrt{12}\left(x^{-\sqrt{12}}-1\right)}.
\end{equation}
In the last one, we have taken a second integration constant in
such a way that $a\rightarrow0$ when $t\rightarrow0$.

Solution (\ref{119}) is plotted in Figure 1. From it we note that
equation (\ref{110}) originates two different universes. In one of
them, we have $a=0$ at $t=0$ and, when $t\rightarrow+\infty$,
$a/A\rightarrow1$ asymptotically. On the other hand, $q=\sqrt{12}-1$
for $a=0$, tending to $+\infty$ as $t\rightarrow+\infty$. Since $q$
is positive, the expansion is decelerated, with its velocity tending
to zero as $a/A\rightarrow1$.

In the second universe, we have the origin of time in $-\infty$,
with $a$ expanding from its asymptotic value $a/A=1$ to $+\infty$,
when $C_1t\rightarrow-1/\sqrt{12}$. On the other hand, $q$ is
initially $-\infty$, when $a/A\rightarrow1$, and increases
approaching $-(\sqrt{12}+1)$ asymptotically, as
$a/A\rightarrow+\infty$. As $q$ is negative, the expansion is always
accelerated.

\begin{figure}
\psfig{figure=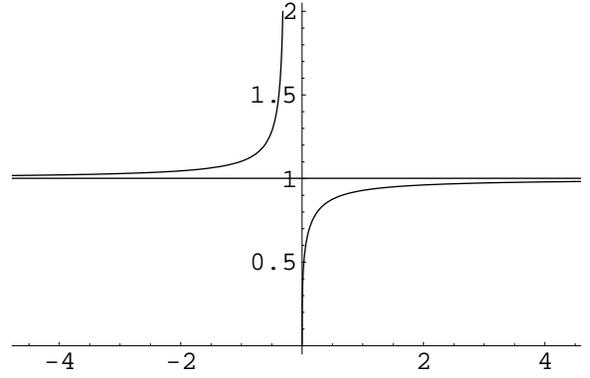}
\caption{Solution (\ref{119}): $a/A$
versus $C_1t$.}
\end{figure}

\vspace{8 mm}

{\emph{2.2 - The solution of equation (\ref{111})}

\vspace{5 mm}

By performing the same transformations and the same steps used to
solve (\ref{110}), we obtain the following equations:
\begin{equation}\label{128}
q=\frac{Cx^{\frac{1}{D}}-B}{x^{\frac{1}{D}}+1},
\end{equation}
\begin{equation}\label{129}
H=\frac{C_1x^{B-1}}{\left(x^{\frac{1}{D}}+1\right)^{D(B+C)}},
\end{equation}and
\begin{equation}\label{130}
t=\int\frac{\left(x^{\frac{1}{D}}+1\right)^{D(B+C)}}{C_1x^B}\,dx,
\end{equation}
with $B$, $C$ and $D$ defined as before.

Equations (\ref{128})-(\ref{130}) are solutions of (\ref{85}) and
(\ref{87}). Let us verify whether they satisfy the third field
equation, (\ref{86}). Once more, performing the same transformations
used to solve (\ref{110}), we arrive at the same equation
(\ref{115}) derived before. As we have seen, only with $\omega=-1$
and $\beta/G_0=-3$ we have the coefficients of $q$, $q^2$ and $q^3$
identically zero, simultaneously. Then, we have again
$B=\sqrt{12}+1$, $C=\sqrt{12}-1$, and $D=-1/\sqrt{12}$.

Therefore, equations (\ref{111}) and (\ref{128})-(\ref{130}) can
be written as
\begin{equation}\label{131}
x=\left(\frac{\sqrt{12}+1+q}{\sqrt{12}-1-q}\right)^{-\frac{1}{\sqrt{12}}},
\end{equation}
\begin{equation}\label{132}
q=\frac{(\sqrt{12}-1)x^{-\sqrt{12}}-\sqrt{12}-1}{x^{-\sqrt{12}}+1},
\end{equation}
\begin{equation}\label{133}
H=C_1 x^{\sqrt{12}}\left(x^{-\sqrt{12}}+1\right)^2,
\end{equation}
\begin{equation}\label{134}
C_1t=-\frac{1}{\sqrt{12}\left(x^{\sqrt{12}}+1\right)}+\frac{1}{\sqrt{12}}.
\end{equation}
In the last one, we have chosen the second integration constant in
such a way that $a\rightarrow0$ when $t\rightarrow0$.

\begin{figure}
\psfig{figure=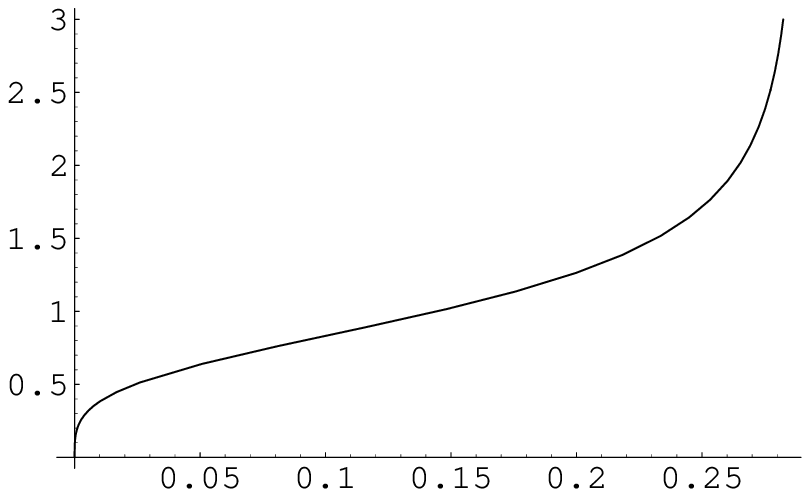}
\caption{Solution (\ref{134}): $a/A$
versus $C_1t$.}
\end{figure}

Solution (\ref{134}) is plotted in Figure 2. We have $a=0$ at $t=0$,
and $a\rightarrow+\infty$ when $C_1t\rightarrow1/\sqrt{12}$. On the
other hand, $q=\sqrt{12}-1$ for $a=0$, decreases with the expansion,
becomes negative, and tends to $-(\sqrt{12}+1)$ when
$a\rightarrow+\infty$.

With the help of (\ref{131}), we can express equations (\ref{133})
and (\ref{134}) as functions of $q$:
\begin{equation}\label{135}
H=-\frac{48C_1}{q^2+2q-11},
\end{equation}
\begin{equation}\label{136}
C_1t=\frac{\sqrt{12}-1-q}{24},
\end{equation}
with $q$ in the interval $(-\sqrt{12}-1,\sqrt{12}-1]$.

The age parameter, on the other hand, can be obtained from
(\ref{135}) and (\ref{136}), leading to
\begin{equation}\label{137}
Ht=\frac{2}{q+\sqrt{12}+1}.
\end{equation}

By using (\ref{82}), (\ref{87}) and (\ref{135}), we obtain
\begin{equation}\label{138}
\rho_m=-24\lambda G_0C_1\left(\frac{q^2+8q+19}{q^2+2q-11}\right).
\end{equation}
The relative density of matter can then be obtained with the help of
equations (\ref{135}) and (\ref{138}), and is given by
\begin{equation}\label{139}
\frac{\Omega_m}{G_0}=\frac{1}{6}(q^2+8q+19).
\end{equation}

Equations (\ref{138}) and (\ref{139}) can be expressed in terms of
$x$, by using (\ref{132}). We have
\begin{equation}\label{140}
\frac{\rho_m}{3\lambda
G_0C_1}=(4-\sqrt{12})x^{\sqrt{12}}+(4+\sqrt{12})x^{-\sqrt{12}}
\end{equation}and
\begin{equation}\label{141}
\frac{\Omega_m}{G_0}=\frac{(4+\sqrt{12})x^{-2\sqrt{12}}+4-\sqrt{12}}
{\left(x^{-\sqrt{12}}+1\right)^2},
\end{equation}
which is plotted in Figure 3.

\begin{figure}
\psfig{figure=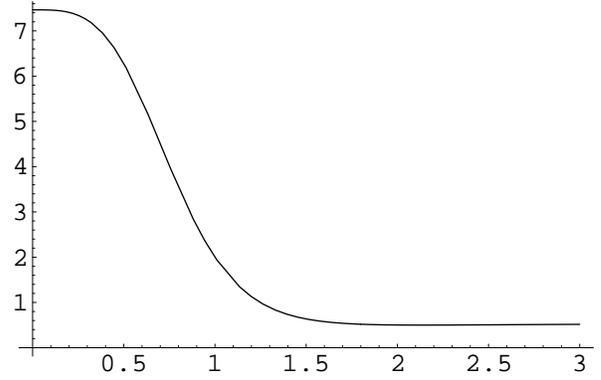}
\caption{Solution (\ref{134}):
$\Omega_m/G_0$ versus $a/A$.}
\end{figure}

\vspace{8 mm}

{\emph{2.3 - The case $\kappa^2<0$}

\vspace{5 mm}

The solutions found in the previous sections were derived from the
differential equation (\ref{88}) by assuming $\kappa^2>0$. Let us
now suppose that it is negative. We can solve equation (\ref{88}) by
doing
\[\kappa_1^2=-\frac{G_0(6\omega+6)^2+6\omega(G_0+\beta)}{G_0\omega^2}\] and
\[z=q+\frac{6+7\omega}{\omega}.\]

In this way, we obtain
\[-\frac{6}{\omega} \int \frac{dz}{\kappa_1^2+z^2}=\frac{3}{3+2\omega}
\int \frac{da}{a},\] which solution is
\[a=C_1 \exp\left(-\frac{6+4\omega}{\omega
\kappa_1}\arctan\frac{q+\frac{6+7\omega}{\omega}}{\kappa_1}\right),\]
where $C_1$ is an integration constant.

Introducing $D_1=\omega\kappa_1/(6+4\omega)$ and
$E=(6+7\omega)/\omega$, one has
\begin{equation}\label{143}
a=C_1 \exp\left(-\frac{1}{D_1}\arctan\frac{q+E}{\kappa_1}\right).
\end{equation}
Taking its inverse function and defining $x=\ln(a/C_1)$, we obtain
\begin{equation}\label{144}
q=-\kappa_1\tan(D_1x)-E,
\end{equation}
or, by using the definition of $q$,
\[\ddot{x}+\dot{x}^2[1-\kappa_1\tan(D_1x)-E]=0.\]

After doing $y=\dot{x}$ and $y'=dy/dx$, we obtain
\[y'+y[1-\kappa_1\tan(D_1x)-E]=0,\]
whose solution is
\begin{equation}\label{145}
y(x)=H=C_2e^{(E-1)x}[\cos(D_1x)]^{-\frac{\kappa_1}{D_1}},
\end{equation}
where $C_2$ is a second integration constant.

As $dt=dx/y$, we also have
\begin{equation}\label{146}
t=\int\frac{[cos(D_1x)]^{\frac{\kappa_1}{D_1}}}{C_2 e^{(E-1)x}}\,dx.
\end{equation}

Equations (\ref{144})-(\ref{146}) are solutions of Brans-Dicke
equations (\ref{85}) and (\ref{87}). As before, let us verify
whether they also satisfy the remaining equation, (\ref{86}).
Introducing $\rho'=d\rho/dx$ and using $\rho$ given by (\ref{87}),
equation (\ref{86}) becomes
\begin{equation}
\label{extra} 3\lambda
G_0\frac{dH}{dx}\left[2+q-\frac{\omega}{6}(1+q)^2\right]\]
\[+\;3\lambda G_0H\frac{dq}{dx}\left[\frac{3-\omega(1+q)}{3}\right]\]
\[+\;3\left\{3\lambda
G_0H\left[1+q-\frac{\omega}{6}(1+q)^2\right]-\beta\lambda
H\right\}=0.
\end{equation}
By using (\ref{145}), one obtains
\[\frac{\beta}{G_0}=\left[E+2+\kappa_1\tan(D_1x)\right]
\left[2+q-\frac{\omega}{6}(1+q)^2\right]\]
\[+\frac{dq}{dx}
\left[\frac{3-\omega(1+q)}{3}\right].\] Now, with the help of
(\ref{144}), we arrive at
\begin{equation}\label{147}
\frac{\beta}{G_0}=(2-q)\left[2+q-\frac{\omega}{6}(1+q)^2\right]+
\frac{\dot{q}}{3H}\left[3-\omega(1+q)\right],
\end{equation}
since $dq/dx=\dot{q}/H$.

On the other hand, from (\ref{88}) one can obtain
\[\frac{\dot{q}}{3H}=\frac{2+q-\frac{\omega}{6}(1+q)^2-(3+2\omega)(1+q)+
\frac{\beta}{G_0}}{3+2\omega}.\] Leading this expression into
(\ref{147}), we have
\[9+\frac{3\omega}{2}+\frac{3\omega^2}{2}-3\omega\frac{\beta}{G_0}+
\left(-6-\frac{11\omega}{2}+\frac{7\omega^2}{2}-\omega
\frac{\beta}{G_0}\right)q\]
\begin{equation}\label{148}
+\left(-3-\frac{\omega}{2}+\frac{5\omega^2}{2}\right)q^2+
\left(\frac{\omega}{2}+\frac{\omega^2}{2}\right)q^3=0.
\end{equation}

It is possible to verify that (\ref{115}) and (\ref{148}) are
identical. In the later, $\omega$ and $\beta/G_0$ are constants.
Therefore, for a varying $q$, the coefficients of $q$, $q^2$ and
$q^3$ must be simultaneously zero, which is only possible if
$\omega=-1$ and $\beta/G_0=-3$. But, in this case, $\kappa_1^2=-12$,
contrary to our initial supposition that $\kappa_1^2>0$.

Therefore, in the case $\dot{q}\neq0$, equation (\ref{143})
satisfies the Brans-Dicke equations (\ref{85}) and (\ref{87}), but
not (\ref{86}). The later is satisfied only if $\dot{q}=0$, in
which case equation (\ref{147}) reduces to (\ref{96}), already
studied.

\vspace{8 mm}

{\emph{2.4 - The case $\kappa^2=0$}

\vspace{5 mm}

In order to fulfill all the possible cases (and solutions) let us
suppose that $\kappa^2=0$. From (\ref{88}) we have
\[-\frac{6}{\omega}\int\frac{dz}{z^2}=\frac{3}{3+2\omega}
\int\frac{da}{a},\]where we have done, as before,
\[z=q+\frac{6+7\omega}{\omega}.\] Its solution is
\[\frac{1}{q+\frac{6+7\omega}{\omega}}=\frac{\omega}{2(3+2\omega)}\ln\frac{a}{C_1},\]
where $C_1$ is an integration constant.

Taking $D_2=\omega/[2(3+2\omega)]$ and, as before,
$E=(6+7\omega)/\omega$, one obtains
\[\frac{1}{q+E}=D_2\ln\frac{a}{C_1}.\]
Introducing the new variable $x=\ln(a/C_1)$, we then have
\begin{equation}\label{149}
q=\frac{1}{D_2x}-E.
\end{equation}
With the definition of $q$, this becomes
\[\dot{x}^2+D_2(1-E)x\dot{x}^2+D_2x\ddot{x}=0,\] or, by doing $F=D_2(1-E)$,
\[\dot{x}^2+Fx\dot{x}^2+D_2x\ddot{x}=0.\]

Taking now $y=\dot{x}$ and $y'=dy/dx$, we obtain
\[y+Fxy+D_2xy'=0,\]which solution is
\begin{equation}\label{150}
y(x)=H=C_2\exp\left({-\frac{Fx+\ln{x}}{D_2}}\right),
\end{equation}
where $C_2$ is another integration constant.

As in the previous case, equations (\ref{149}) and (\ref{150}) are
solutions of (\ref{85}) and (\ref{87}), but we should also verify
whether they satisfy (\ref{86}). Introducing $\rho'=d\rho/dx$ and
using $\rho$ given by (\ref{87}), equation (\ref{86}) reduces, as we
have seen, to (\ref{extra}). Now, using equations (\ref{150}) and
(\ref{149}) leads to the same equation (\ref{147}) of the previous
case, which, as already seen, leads to (\ref{148}). Therefore,
solutions with varying $q$ are only possible if $\omega=-1$ and
$\beta/G_0=-3$. With these values, however, $\kappa^2=12$, contrary
to our initial supposition that $\kappa^2=0$. We thus conclude, also
in this case, that (\ref{86}) is not satisfied, that is, there is no
solution with varying $q$.

\section{Conclusions}

In this work we have found some exact solutions of Brans-Dicke
cosmology, by using two different ansatzen. We have shown that the
first ansatz is a particular case of the second one, when the
deceleration parameter $q$ is constant.

In the first ansatz, the ratio between the energy densities of
matter and vacuum is constant, characterizing a possible solution
for the cosmic coincidence problem, that is, the approximated
coincidence presently observed between $\rho_m$ and $\rho_c$. This
possibility survives to a quantitative analysis, since a relative
matter density around $1/3$, as indicated by observations, leads
to an age parameter $Ht\approx1$, corresponding to a universe age
around $14$ Gyr, also in accordance with observational limits.

Nevertheless, this ansatz presents some problems as well. The most
severe of them is the presence of a constant deceleration factor
(equals to zero if $Ht = 1$). In spite of the claim of some authors
(see, for example, \cite{jain,sethi}) about the possibility of a
uniform expansion along the whole universe evolution, a decelerated
phase is usually considered necessary for large structure formation.
For this reason, we should consider this ansatz valid only in the
limit of late times, restricting in this way the predictive power of
the model.

With the second ansatz, besides the case of constant $q$, we have
found three other universes, with varying $q$, in which the dark
energy density is negative and the Brans-Dicke parameter is
$\omega = -1$. In one of them the deceleration parameter is always
highly positive. In a second one, it is always highly negative.
Therefore, these two cases are interesting just from a theoretical
viewpoint.

In the third case, on the other hand, the deceleration parameter is
initially positive, becoming negative at later times, but always
finite. In this case (as well as in one of the previous cases) one
has a future {\it big-rip}, with the scale factor, the matter
density and the Hubble parameter diverging in a finite time, but
with the relative matter density remaining finite. As one can see
from equation (\ref{137}), for an age parameter in the interval
$0.8<Ht<1.3$, as defined by the observational limits, the
deceleration parameter is $-2.0>q>-2.9$. Whence, with the help of
(\ref{139}), it is possible to see that we have, for the relative
matter density, $1.1>\Omega_m/G_0>0.7$. As we know, different
observations restrict the matter density parameter to
$0.2<\Omega_m<0.4$. Therefore, this solution satisfies such
observations, provided $G_0$ is in the interval $0.3<G_0<0.4$.
Furthermore, for the whole evolution we have $\Omega_m/G_0<7.5$,
tending, for future times, to a constant value around $0.5$ (see
Figure 3). This also characterizes a possible explanation for the
cosmic coincidence.

It is interesting to observe that, in the three cases with varying
$q$, the cosmological time varies linearly with $q$, which may,
therefore, be used to define the time measurement. It is
interesting to note as well that the total energy density can be
negative, since the dark energy density is negative. In our last
solution, for example, $\rho$ is positive until $q\simeq-2.3$,
becoming negative since then.

There is a particular result which may seem a limitation of our
solutions, namely the typical values found for the Brans-Dicke
parameter $\omega$. As we know, observations in the realm of Solar
System impose very high inferior limits for it. Let us remember,
however, that we are considering the simplest version of
scalar-tensor theories, which plays just an effective role here.
Corrections to General Relativity, if exist, may be scale-dependent,
and, therefore, observations in the Solar System cannot, in
principle, impose limits to corrections at the cosmological scale.
Particularly, we should not expect any time dependence of the
Brans-Dicke scalar field (and so of $G$) in the Solar System, where
the metric is stationary. While no spatial dependence should exist
in large scale, because of the cosmological principle.

Anyway, a generalization of the solutions studied here seems to be
necessary, either by modifying our ansatzen in the case of early
times, either by considering more general scalar-tensor theories,
with $\omega$ depending on the scale. The study performed here,
though limited in its scope, shows the variability of solutions in
these contexts.

We should also note that, for the solution of Brans-Dicke equations,
it is enough to add, for instance, the Eddington-Dirac relation. The
inclusion of additional constraints, even when empirically or
theoretically justified, had the goal of limiting the set of
solutions. Therefore, a possible line of investigation would be to
relax our ansatzen, imposing to the Brans-Dicke equations, for
example, just the constraint given by the Eddington-Dirac relation,
enlarging in this way the class of possible solutions. Such a
generalization may also include the radiation phase, although we do
not have any indication about the validity of the Eddington-Dirac
relation at early times.

To conclude, let us remind that, despite the analysis we have made
about the observational limits for the age and matter density
parameters, a more detailed analysis of the whole set of current
observational data is still in order. In particular, a careful
study of the distance-redshift relation for supernovas Ia
constitutes the subject of a forthcoming publication.

\section*{Acknowledgments}

The authors are grateful to J. C. Fabris and R. Muniz for helpful
comments, and to the anonymous referees for constructive
suggestions. SC is partially supported by CNPq (Brazil).

\end{document}